
\hsize=5.6 true in
\vsize=9.1 true in
\hoffset=0.24 true in
\parindent=20pt
\parskip=2.2pt
\baselineskip=16true pt plus.2 true pt
\font\fourbf=cmbx10 scaled 1440
\font\fourrm=cmr10 scaled 1440
\font\twelverm=cmr12
\font\twelveit=cmti12
\font\twelvebf=cmbx12
\font\twelvei=cmmi12
\font\twelvesy=cmsy10 scaled 1200

\font\tenbf=cmbx10 scaled 1097
\font\ninebf=cmbx9
\font\ninerm=cmr9
\font\ninei=cmmi9
\font\ninesy=cmsy9
\font\ninesy=cmsy9
\font\seveni=cmmi7
\font\sevenrm=cmr7
\font\sevensy=cmsy7
\font\sevenbf=cmbx7

\def\twelvepoint{\def\rm{\fam0\twelverm}
    \textfont\itfam=\twelveit \textfont\bffam=\twelvebf
      \textfont0=\twelverm \scriptfont0=\ninerm   \scriptscriptfont0=\sevenrm
      \textfont1=\twelvei  \scriptfont1=\ninei  \scriptscriptfont1=\seveni
      \textfont2=\twelvesy \scriptfont2=\ninesy  \scriptscriptfont2=\sevensy
        \def\bf{\fam\bffam\twelvebf}
        \def\it{\fam\itfam\twelveit}
        \textfont\bffam=\twelvebf
        \scriptfont\bffam=\ninebf \scriptscriptfont\bffam=\sevenbf
   \rm}
\def\w{\widetilde}
\def\et{{\it et al.}}

\def\J{{\cal J}}
\def\H{{\cal H}}
\def\R{{\bf R}}
\def\q{{\bf q}}
\def\0{{\bf 0}}
\def\A{\overline}
\def\B{\A{\A{\chi}}}
\def\C{\B_{}^{\,o}}
\def\Cr{\B_r^{\,o}(\omega)}
\def\la{\langle}
\def\ra{\rangle}
\def\half{{\textstyle{1\over2}}}
\def\wJ{\w{J\kern2pt}\kern-2pt}
\def\us{\hskip 1.8pt}
\hbox{}
\vskip 0.1 true in
{\twelvepoint
\centerline{\fourbf 1/z-renormalization of the mean-field behavior of}
\medskip
\centerline{\fourbf the dipole-coupled singlet--singlet system
HoF$_{\hbox{\tenbf3}}$}
\bigskip
\bigskip
\bigskip
\bigskip
\centerline{\fourrm Jens Jensen}
\medskip
\centerline{\rm Niels Bohr Institute, \O rsted Laboratory,
Universitetsparken 5,}
\centerline{\rm 2100 Copenhagen, Denmark}
\vfill
\bigskip
\noindent
\centerline{\bf ABSTRACT}
\medskip
\noindent
The two main characteristics of the holmium ions in HoF$_3$ are that
their local electronic properties are dominated by two singlet states
lying well below the remaining 4f-levels, and that the classical
dipole-coupling is an order of magnitude larger than any other two-ion
interactions between the Ho-moments. This combination makes the
system particularly suitable for testing refinements of the mean-field
theory. There are four Ho-ions per unit cell and the hyperfine coupled
electronic and nuclear moments on the Ho-ions order in a
ferrimagnetic structure at $T_C=0.53$\us K.
The corrections to the mean-field behavior of holmium triflouride,
both in the paramagnetic and ferrimagnetic phase, have been calculated
to first order in the high-density $1/z$-expansion. The effective
medium theory, which includes the effects of the single-site
fluctuations, leads to a substantially improved description of the
magnetic properties of HoF$_3$, in comparison with that based
on the mean-field approximation.
\footnote{}{PACS-numbers: 75.10.-b \ ; \ 75.30.-m \ ; \ 75.50.Gg}
\bigskip
\bigskip
\bigskip
\bigskip
\eject
\hbox{}
\medskip
\noindent
{\bf I. Introduction}
\medskip
\noindent
The magnetic properties of HoF$_3$ have been established
experimentally$^{1-3}$ in considerable detail, and these experiments were
recently interpreted in terms of a mean-field (MF) model by Leask
\et$^1$ -- HoF$_3$ is orthorhombic, and the low symmetry at the Ho
sites implies that the $J=8$
ground-state multiplet splits into singlets. In the paramagnetic phase,
the energy difference between the two lowest singlets is about 0.7\us
meV and the dipole matrix-element between the two states is large. The
next singlet lies at about 5\us meV whereas the remaining ones lie
between 10 and 50\us meV above the ground state,$^4$ and this arrangement
of the crystal-field levels leads to a very anisotropic susceptibility
at low temperatures. The system is close to an Ising system, with
the only modification that there are two easy directions, one for each
magnetically-equivalent pair of Ho-ions in a unit cell.
The classical dipole-coupling between
the angular moments is weak, but is nevertheless found to be one order
of magnitude stronger than any other interaction between the dipoles.
It gives rise to strong correlation effects below 5\us{K} and is
responsible for the induced magnetic ordering of the singlet
ground-state system  at $T_C=0.53$\us K. The dipole coupling is
however not sufficiently strong to produce an ordering of the isolated
electronic system. The ratio between the two-ion coupling and the
threshold value required for inducing magnetic ordering of the
electronic moments is found to be $R=0.86$. The phase transition
occurs only because the hyperfine interaction between the electronic and
nuclear moments enhances the effective susceptibility, thus leading to a
cooperative ordering of the two systems. Below $T_C$ the ordered
moments are along the two easy directions.
At low temperatures, HoF$_3$ may be considered to be a simple
singlet--singlet system in which the moments interact like classical
dipoles. The dipole-dipole interaction can be calculated accurately
from first principles and the magnetic system is fully characterized
by only a few parameters. The only complications are that there
are four magnetic ions per unit cell, that the hyperfine
interaction plays an active role, and that additional two-ion
couplings are of some importance.

Although the mean-field theory of Leask \et$^1$ was able to reproduce many
of the observations made in HoF$_3$, such as the excitation
spectrum determined by neutron scattering at 1.6\us K, some
discrepancies remained. The calculated moment in the zero temperature
limit was 16\% smaller than that observed, and the comparison
between the theoretical and experimental heat capacity was not
entirely acceptable. Quadrupole--quadrupole couplings may be of some
importance in HoF$_3$ for explaining the discrepancies, but in this
paper we shall concentrate on the corrections to the mean-field theory
due to the single-site fluctuations. In section II we recapitulate
the results of the high-density $1/z$-expansion for the
singlet--singlet model utilizing the effective medium approach.$^5$
The theory is then extended to cover both the paramagnetic and the
ordered case. In section III the theory is applied to HoF$_3$ and the
conclusions are presented in the final section IV.

\bigskip
\noindent
{\bf II. The 1/z-expansion}
\medskip
\noindent
We consider a Bravais lattice of $N$ identical singlet--singlet
atoms, characterized by the energy separation $\Delta$ between the two
singlets, and by the dipole matrix-elements. In the case of a
singlet--singlet system, the $x$-axis may be defined so that only the
matrix-elements of $J_x^{}$ are non-zero, and $M$ denotes the
(numerical) value of this $J_x^{}$ matrix-element between the two
states, whereas $+m$ and $-m$ are the matrix-elements of $J_x^{}$
within respectively the lower and upper singlet. $M_0^2=M^2+m^2$ is a
constant and $m=0$ in the paramagnetic case at zero field. At the
temperature $T=1/k_B^{}\beta$, the MF population-factors of the lower
and the upper state are $n_0^{}$ and $n_1^{}$, where
$n_0^{}=1/[1+\exp(-\beta\Delta)]$, $n_0^{}+n_1^{}=1$ and we define
$n_{01}^{}=n_0^{}-n_1^{}$.

The two-site Green function is defined as the $\tau$-ordered ensemble
average
$$
G(ij,\tau_1^{}-\tau_2^{})=-\la T_\tau^{}\wJ_{ix}^{}(\tau_1^{})\wJ_{jx}^{}
(\tau_2^{})\ra\eqno(2.1)
$$
$|\tau_1^{}-\tau_2^{}|\le \beta$, and we use the short-hand notation
$$
\wJ_{ix}^{}=J_{ix}^{}-\la J_{x}^{}\ra\eqno(2.2)
$$
The Fourier transform of the Green function is defined
in terms of ($\hbar$ times) the Matsubara frequencies, $\omega_n^{}=2\pi
n/\beta$, where $n$ is an integer:
$$
G(\q,i\omega_n^{})=\sum_{j}\,e^{-i\q\cdot\R_{ij}}\int_0^\beta
G(ij,\tau)\,e^{i\omega_n^{}\tau}\,d\tau\eqno(2.3)
$$
The Hamiltonian is divided into two parts, $\H=\H_0^{}+\H_1^{}$, where
$\H_0^{}$ is the mean-field part and $\H_1^{}$ is the perturbation
$$
\H_1^{}=-\half\sum_{ij}\J(ij)\wJ_{ix}^{}\wJ_{jx}^{} \eqno(2.4)
$$
in which case the Green function is determined by the linked-cluster
expansion
$$
G(ij,\tau)=-{{\la T_\tau^{}\,U(\beta,0)\wJ_{ix}(\tau)\wJ_{jx}(0)\ra_0^{}}
\over{\la U(\beta,0)\ra_0^{}}}\eqno(2.5)
$$
with
$$
U(\beta,0)=1-\int_0^\beta\H_1^{}(\tau_1^{})d\tau_1^{}+\cdots\eqno(2.6)
$$
The index 0 on the thermal averages indicates that they are the mean-field
values. The non-interacting Green function, obtained
in the zeroth order of $\H_1^{}$, vanishes if $i\ne j$, whereas if
$i=j$ it is
$$
G_0^{}(i\omega_n^{})=-M^2g(i\omega_n^{})-m^2h(i\omega_n^{})\eqno(2.7)
$$
where the two response functions are
$$
g(i\omega_n^{})={{2n_{01}^{}\Delta}\over{\Delta^2-(i\omega_n^{})^2}}\,;
\quad h(i\omega_n^{})=\beta(1-n_{01}^2)\delta_{n,0}^{}\eqno(2.8)
$$

The perturbation $\H_1^{}$ cannot, in general, be considered as being
small compared to $\H_0^{}$, but each time a term involving the
two-ion coupling is summed over $\q$, we effectively gain a factor
$1/z$, where $z$ is the co-ordination number. A systematic expansion
of the Green function in powers of $1/z$, in the case of the
singlet--singlet system, was developed by Stinchcombe.$^6$ Here, we
shall use a slightly different approach and utilize the concept of an
effective medium, which is the basis of the coherent-potential
approximation.

The ensemble averages in (2.5) are calculated by expressing the angular
momentum operators in terms of the `standard-basis' operators,
$a_{\nu\mu}^{}=|\nu\ra\la\mu|$ where $|\nu\ra$ are the MF-eigenstates
of an atom, and by utilizing the invariance of the trace to a cyclic
permutation of the operators.$^{7,8}$ These are not Bose-operators, so
the `contractions' determined by the commutators of the different
operators are not $c$-numbers, but operators which give rise to new
contractions. However, the next generation of contractions adds terms
to $G(\q,i\omega_n^{})$ which always involve additional $\q$-summations.
Hence in the order $(1/z)_{}^0$, these contractions are neglected.
This corresponds to a decoupling of the higher-order cumulants in (2.5)
into products of the second-order terms, $\la T_\tau^{}\wJ_{ix}^{}(\tau_1^{})
\wJ_{jx}(\tau_2^{})\ra_0^{}=-\delta_{ij}^{}G_0^{}(\tau_1^{}-\tau_2^{})$
(if the possible difference between $\la J_x^{}\ra$ and $\la
J_x^{}\ra_0^{}$ is neglected, see below). The infinite series of
`chain diagrams', generated by (2.5) using this decoupling, is easily
summed, and the result is:
$$
G(\q,i\omega_n^{})\Big|_{\rm RPA}={G_0^{}(i\omega_n^{})
\over{1+\J(\q)G_0^{}(i\omega_n^{})}} \eqno(2.9)
$$
showing that this approximation is equivalent to the random-phase
approximation. The difference between the fourth-order cumulant
and the corresponding decoupled value appears in the next order of
$1/z$. In the usual `unconditional' cumulant expansion this
difference, the fourth-order semi-invariant, is introduced as an
additional vertex. The vertex in the RPA chain-diagrams is replaced by
the sum of this and the fourth-order semi-invariant, and neglecting
any particular effects of whether some vertices in a chain belong to
the same site or not, we may straightforwardly sum the series.$^6$

In the order $1/z$ only the single-site Green function is directly
modified, as the fourth-order cumulant only differs from the decoupled
one if all the four momentum operators belong to the same site. The
single-site Green function is
$$
G(i\omega_n^{})\equiv G(ii,i\omega_n^{})={1\over N}
\sum_{\q}G(\q,i\omega_n^{})\eqno(2.10)
$$
and, to the order $1/z$, the two-site Green function may be expressed
in terms of $G(i\omega_n^{})$ by introducing the effective-medium
coupling
$$
K(i\omega_n^{})={1\over N}\sum_{\q}
\J(\q)G(\q,i\omega_n^{})/G(i\omega_n^{})\eqno(2.11)
$$
in which case
$$
G(\q,i\omega_n^{})={G(i\omega_n^{})\over{1+\{\J(\q)-K(i\omega_n^{})\}
G(i\omega_n^{})}}\eqno(2.12)
$$
$K(i\omega_n^{})$ is the sum of all chain diagrams which start and
end at the same site without crossing this site in between.$^5$ We
may therefore consider an effective cumulant expansion of
$G(i\omega_n^{})$, equivalent to (2.5) where $\J(ij)$ in $\H_1^{}$ is
replaced by the time-dependent coupling $K(\tau_1^{}-\tau_2^{})$, and
the term to leading order in this coupling is
$$
\eqalign{
-&\half\int_0^\beta\!\!\int_0^\beta\!\!\int_0^\beta d\tau
d\tau_1^{}d\tau_2^{}\,e^{i\omega_n^{}\tau}
{1\over\beta}\sum_{n'}K(i\omega_{n'}^{})\,e^{i\omega_{n'}^{}
(\tau_1^{}-\tau_2^{})}\cr &{}\times
\Big[\la T_\tau^{}\wJ_{ix}^{}(\tau_2^{})
\wJ_{ix}(\tau_1^{})\wJ_{ix}^{}(\tau)\wJ_{ix}(0)\ra_0^{}
-\la T_\tau^{}\wJ_{ix}^{}(\tau_2^{})
\wJ_{ix}(\tau_1^{})\ra_0^{}\la
T_\tau^{}\wJ_{ix}^{}(\tau)\wJ_{ix}(0)\ra_0^{}\Big]}\eqno(2.13)
$$
The decoupling of the higher-order cumulants in the single-site series
into products of the second-order terms leads to the result
$$
G(i\omega_n^{})\Big|_{\rm RPA}=
{G_0^{}(i\omega_n^{})\over{1+K(i\omega_n^{})G_0^{}(i\omega_n^{})}}\eqno(2.14)
$$
and when this result is introduced in (2.12), $K(i\omega_n^{})$ cancels out
and we get the RPA-result, (2.9). To the next order in $1/z$
we have to include the fourth-order cumulant term, (2.13). We shall
first consider the paramagnetic phase, $m=0$, in which case
$$
G(i\omega_n^{})=G_0^{}(i\omega_n^{})-G_0^{}(i\omega_n^{})
\{K(i\omega_n^{})G_0^{}(i\omega_n^{})+\Sigma(i\omega_n^{})\}+\cdots
\eqno(2.15)
$$
where the renormalization factor is
$$
\Sigma(i\omega_n^{})=\alpha+\gamma(i\omega_n^{})g(i\omega_n^{})\eqno(2.16)
$$
$\alpha$ is a constant
$$
\alpha={M^2\over n_{01}^2}\Big[\lambda_2^{}-\half
\big\{g(0)+\beta(1-n_{01}^2)\big\}\lambda_1^{}\Big]\eqno(2.17)
$$
and the frequency-dependent term is
$$
\gamma(i\omega_n^{})={M^2\over
n_{01}^2}\Big[\lambda_1^{}-(1-n_{01}^2)K(i\omega_n^{})\Big]\eqno(2.18)
$$
and the parameters $\lambda_p^{}$ are defined as
$$
\lambda_p^{}={1\over\beta}\sum_{n'}
K(i\omega_{n'}^{})\big[g(i\omega_{n'}^{})\big]_{}^p\eqno(2.19)
$$
The unconditional cumulant expansion accounts correctly for the
fourth-order cumulant term in (2.15), but an analysis of the sixth and
higher-order terms shows that this procedure does not lead to a good
estimate of the higher-order contributions in the single-site series.
Instead, it is found that the series generated by replacing
$G_0^{}(i\omega_n^{})$ in front of the second term  of the single-site
series (2.15) by the interacting Green function $G(i\omega_n^{})$,
much more effectively accounts for the terms deriving from,
specifically, the sixth-order cumulant. Introducing this Dyson-like
result for the single-site Green function in (2.12), we get:
$$
G(\q,i\omega_n^{})=
{G_0^{}(i\omega_n^{})\over{1+\Sigma(i\omega_n^{})+
\J(\q)G_0^{}(i\omega_n^{})}}\eqno(2.20)
$$
valid to first order in $1/z$. This
result is nearly the same as the one derived by Galili and Zevin$^9$
using an elaborate renormalization of the unconditional expansion. In
addition to the simplifications attained by utilizing the
effective-medium approach, the present procedure is fully
self-consistent. A more detailed discussion of the effective
medium theory and its comparison with the unconditional cumulant
expansion may be found in Refs.\ 5 and 10. In a systematic
expansion in $1/z$, the effective-medium approximation ceases to be
valid in second order. However, an improvement of the theory
is obtainable by including the new diagrams due to the sixth-order
cumulants into the single-site series, neglecting the corresponding
$(1/z)^2$ two-site effects. The second-order contributions to the
effective medium have
been considered in an analysis$^{11}$ of Pr and are of some importance if
the energy gap in the excitation spectrum is small. In the present analysis
we shall consider only the leading-order $1/z$-modifications due to the
single-site fluctuations.

In the paramagnetic phase, the contribution of the
fluctuations to the internal energy is
$$
\delta U={N\over{2\beta}}\sum_n\Big\{K(i\omega_n^{})-{{\Delta+
i\omega_n^{}}\over M^2}\Big\}G(i\omega_n^{})-Nn_1^{}\Delta \eqno(2.21)
$$
and introducing the $1/z$-result for $G(i\omega_n^{})$ in this
expression, we may write the energy change:
$$
\delta U={N\over2}\Big[{{\alpha n_{01}^{}}\over{1+\alpha}}\Delta-{}
M^2\lambda_1^{}+{1\over{\beta}}\sum_n K(i\omega_n^{})
\big\{G(i\omega_n^{})-G_0^{}(i\omega_n^{})\big\}
\Big]\eqno(2.22)
$$
where the last sum is a small second-order term. The derivative of
$\delta U$ with respect to $T$ determines the change of the heat
capacity from its mean-field value. In the calculation of the energy
it is of importance that the Green function satisfies the sum rule:
$$
{1\over\beta}\sum_n G(i\omega_n^{})=-M^2\eqno(2.23)
$$
($m=0$). This is strictly the case if the single-site series is
terminated after the first or the second term in (2.15). It is also found
numerically to be valid with a rather high accuracy for the
self-consistent $1/z$-Green function.

Next we wish to consider the situation when $m\ne0$. In the
low-temperature regime, with which we shall mostly be concerned,
the elastic pole in (2.7) is weak compared to the inelastic one, unless
the system is close to the saturation limit $m\gg M$. However, in this
limit, i.e.\ at high fields, all the fluctuations are quenched and the
mean-field approximation is valid. This means that, generally, at low
temperature the elastic fluctuations are of less importance than the
inelastic ones, and it is therefore acceptable to include the
elastic-pole contributions less rigorously.

The first problem we meet in the order $1/z$ is that, if we define the
mean-field Hamiltonian in the usual way, $\la \wJ_x^{}\ra_0^{}$ will in
general be non-zero, giving rise to additional complications in the
linked-cluster expansion of the Green function. In order to avoid these
complications we introduce a modified mean field,
$H_{\rm MF}^{}$, defined by the requirement that
$$
\la J_x^{}\ra =\la J_x^{}\ra_0^{}=m^2 n_{01}^{}\eqno(2.24)
$$
which differs from the usual mean field $H_0^{}$:
$$
H_{\rm MF}^{}=H_0^{}-\delta H\, ;
\quad g\mu_B^{}H_0^{}=g\mu_B^{}H+\J(\0)\la J_x^{}\ra\eqno(2.25)
$$
where $H$ is the applied field. The determination of $\H_0^{}$ in terms of
$H_{\rm MF}^{}$ instead of $H_0^{}$  introduces an
extra perturbation in
the Hamiltonian, $\H=\H_0^{}+\H_1^{}+\H_2^{}$, with $\H_2^{}=
-g\mu_B^{}\delta H\sum_i \wJ_{xi}^{}$. The ratio
$\delta H/H_{\rm MF}^{}$ is a small quantity, at the most about $0.1$
in the numerical calculations, and $\H_2^{}$ is truly a weak
perturbation. Furthermore, the leading order modification of the
cumulant expansion is proportional to $(\delta H/H_{\rm MF}^{})^2$,
which small correction is neglected.

The elastic pole disturbs the inelastic fluctuations, and
$\Sigma(i\omega_n^{})$ in (2.20), when $n\ne 0$, is changed into
$$
\Sigma(i\omega_n^{})=\alpha-\alpha_m^{}+\Big[\gamma(i\omega_n^{})-
{{2m^2}\over M^2}\gamma(0)\Big]g(i\omega_n^{})\eqno(2.26)
$$
with
$$
\alpha_m^{}={m^2\over n_{01}^2}\Big[\lambda_2^{}-
g(0)\lambda_1^{}+{4\over{g(0)}}\lambda_3^{}-(1-n_{01}^2)
(1+\half\beta\Delta n_{01}^{})K(0)g(0)\Big] \eqno(2.27)
$$
In this result we have for simplicity neglected some
frequency-dependent terms which vanish in the zero-frequency limit
($g(i\omega_{n'}^{}\pm i\omega_n^{})$ appearing as a factor in the
$n'$-sums has been replaced by its zero-frequency value
$g(i\omega_{n'}^{})$). The inelastic broadening of the elastic peak
discussed in  Ref.11 is neglected in this approximation.
In addition to the inelastic modifications we get elastic terms
appearing only at zero frequency, and at this frequency
the single-site Green function is found to be
$$
G(0)=-{{M^2g(0)}\over{1+\Sigma(0)-M^2K(0)g(0)}}-m^2\xi h(0)\eqno(2.28)
$$
where $\Sigma(0)$ is the zero-frequency value of
$\Sigma(i\omega_n^{})$ given by (2.26) and
$$
\xi=
{{1+\tanh(m^2n_{01}^2\beta K(0)-M^2\beta \lambda_1^{})}\over
{1+\{4n_{01}^2K(0)g(0)+2\lambda_2^{}+g(0)
\lambda_1^{}\}M^2/n_{01}^2}} \eqno(2.29)
$$
The cumulant expansion of the elastic contribution effectively becomes
an expansion in $\beta K(0)$ which converges slowly at low temperatures,
where however the elastic term as a whole is frozen out. The actual
result for the $1/z$-term in the numerator of $\xi$, deriving from the
fourth-order cumulant, is the leading order term in the Taylor
expansion of the tanh-term, which diverges in the zero-temperature
limit. This divergence was not found to influence the numerical
results significantly, as the term is multiplied by $h(0)$ which vanishes
exponentially in this limit. However, in order to account in a
simple way for the renormalization of this term expected due to
higher-order contributions, we have replaced the $1/z$-term by its
tanh-value.

The final result for $G(\q,i\omega_n^{})$ to first order in $1/z$ when
$m$ is non-zero is determined in terms of $G(i\omega_n^{})$. At
non-zero frequencies the result may be written as (2.20), with
$\Sigma(i\omega_n^{})$ now given by (2.26). At zero frequency we have
to introduce $G(0)$ given by (2.28) in the original expression (2.12)
for $G(\q,i\omega_n^{})$.
The result (2.12) becomes equivalent to the RPA-result if
$G_0^{}(i\omega_n^{})$ in (2.9) is replaced by the effective
non-interacting Green function
$$
\w{G}_0^{}(i\omega_n^{})= {{G(i\omega_n^{})}\over{1-K(i\omega_n^{})
G(i\omega_n^{})}} \eqno(2.30)
$$
which is equal to $G_0^{}(i\omega_n^{})/\{1+\Sigma(i\omega_n^{})\}$,
except in the zero-frequency case when $m\ne0$.

The mean field $H_{\rm MF}^{}$ is determined by
(2.24) which may be written
$$
\la J_x^{}\ra =\la J_x^{}\ra_0^{}=-g\mu_B^{}\int_0^{H_{\rm MF}^{}}
G_0^{}(0;H')dH' \eqno(2.31)
$$
where $G_0^{}(0;H')$ is the non-interacting Green function,
(2.7), as a function of the Zeeman-field $H'$. Differentiating this
equation with respect to the applied field $g\mu_B^{}H$, at a constant
temperature, we get on the left-hand side the static
bulk-susceptibility $-G(\0,0)$, or
$$
G(\0,0;H)=G_0^{}(0;H_{\rm MF}^{})\{1-\J(\0)G(\0,0;H)\}\,
{{dH_{\rm MF}}^{}\over{dH_0^{}}}\eqno(2.32)
$$
which is directly integrable, as $H_{\rm MF}^{}$ is the mean field
determining $G(\0,0;H)$, and we get the following relation between
the two field quantities
$$
H_0^{}=\int_0^{H_{\rm MF}^{}}{{G_0^{}(0;H')}\over
{\w{G}_0^{}(0;H')}}\,dH'\eqno(2.33)
$$
making use of the effective non-interacting Green function defined
by (2.30). The combination of (2.33) and (2.25) determines the relation
between the applied field and the mean field, which allows a fully
self-consistent calculation of $\la J_x^{}\ra$ as a function of field,
in much the same way as in the MF-approximation. The adjustment
of the mean field, by replacing $H_0^{}$ by the effective value
$H_{\rm MF}^{}$, implies that the change in the free energy including
the $1/z$-contributions, relatively to the non-magnetic state, is
determined by the mean-field part of the Hamiltonian, ${\cal H}_0^{}$,
because $dF/N=-g\mu_B^{}\la J_x^{}\ra dH$ is equal to
$-g\mu_B^{}\la J_x^{}\ra_0^{} dH$ (at constant temperature). This has
the consequence that
$$
\delta F(m=0)=-N\int_0^{M_0^{}}g\mu_B^{}\delta H \,d\la J_x^{}\ra
\eqno(2.34)
$$
$M_0^{}$ is the saturation value of $\la J_x^{}\ra$ in the
limit of an infinite field, in which limit any correction to the
MF-approximation vanishes. The difference $\delta H=H_0^{}-H_{
\rm MF}^{}$ is considered to be a function of $\la
J_x^{}\ra$, and
introducing $\delta F(m=0)$, as determined by $\delta U$ in (2.22) and
the corresponding change in heat capacity, the fulfillment of (2.34)
provides an independent test of the theory.

\bigskip
\noindent
{\bf III. The magnetic properties of HoF$_{\bf 3}$}
\medskip
\noindent
HoF$_3$ is orthorhombic with the lattice parameters $a=6.404$\us
\AA$^{-1}$, $b=6.875$\us \AA$^{-1}$ and $c=4.379$\us \AA$^{-1}$, and
there are 4 Ho$^{3+}$ and 12 F$^-$ ions per unit cell.
The positions of the ions within the unit cell are
specified in the previous papers$^{2,3}$ on HoF$_3$, and the
projections of the ions on the $a$--$c$ and $a$--$b$ planes are shown in
Fig.1. The crystal-field
Hamiltonian for the Ho-ions at the four different sites in the unit
cell is the same when referring to one local coordinate system for
the sublattices labeled 1 and 2 in Fig.1 and another, in which the
$b$- and $c$-axes are reversed, for the sublattices 3 and 4. The local
easy axes, or $x$-axes, are indicated in the figure. They lie in the
$a$--$c$ plane and make the angles $\theta$ and $-\theta$ with the
$a$-axis. The magnitude of $\theta$, but not its sign, is known from
the experiments.$^{1,2}$ The sign of $\theta$ has not much influence on
the analysis, but we have here chosen $\theta=+25^\circ$ as also used
in Fig.1. This choice of sign leads to a slightly better fit of the
magnetization curves than the other, but the difference is not so
significant that it rules out the alternative choice.

The equivalence of the ions, when considered in their local coordinate
systems, makes it straightforward to generalize the theory in the
preceding section to this case with four magnetic ions per unit cell.
In the paramagnetic phase the only modification of the
MF/RPA-calculations presented in Ref.1 are that the components in
the single-ion susceptibility tensors $\Cr$, deriving from the
singlet--singlet transition on the $r$th ion, should be divided by the
common renormalization factor $1+\Sigma(\omega)$. This follows from
the fact that the RPA-expression for the final Green function is
unchanged if the non-interacting Green function $G_0^{}(i\omega_n^{})$
in (2.9) is replaced by the effective $\w{G}_0^{}(i\omega_n^{})$
given by (2.30). Neglecting small $1/z$-corrections of the
RPA-contributions from other single-ion transitions, this procedure
allows the full level scheme of the ions to be included in the
calculations. The imaginary part of $\Sigma(\omega)$ is
proportional to the imaginary part of $K(\omega)$ times
$(1-n_{01}^2)$, and the latter factor is vanishing small at low
temperatures, thus $\Sigma(\omega)$ is real just above $T_C$.
In this limit the paramagnetic density of states, for each of the $4N$
magnetic ions, is
$$
{\cal N}(\varepsilon)={1\over{4 N}}\sum_{{\bf q},\nu}
\delta(\varepsilon_{\bf q,\nu}^{}
-\varepsilon)\eqno(3.1)
$$
where $\varepsilon_{\bf q,\nu}^{}$ are the four excitation energies
at a certain $\q$. The $\q$-summation
in the expression (2.11) for $K(i\omega_n^{})$ may be replaced by
an energy integration by the introduction of ${\cal N}(\varepsilon)$. The
resulting integral is the same as the one derived for a
Bravais-lattice. Calculating $K(i\omega_n^{})$ in this way, the final
density of states has been determined self-consistently by an
iterative procedure, using the RPA as a starting point. The result, nominally
at 0.55\us K, is shown in Fig.2. The corresponding excitation energies
at 1.6\us K along the high-symmetry directions $(h\,0\,0)$ and
$(0\,0\,l)$ are shown in Fig.3. As discussed in Ref.1 an effective
double-zone representation is applicable along these directions, and
only two modes scatter the neutrons in a constant-$\q$ scan. The calculated
dispersion relations, the solid lines in Fig.3, are compared with the
experimental neutron-scattering results and the RPA-predictions
presented in Ref.1. Although the excitation energies in the order $1/z$
are not changed much compared to the RPA-theory, the tendency
of the excitations to split into two separated bands is much more
pronounced in the final density of states in Fig.2 than in the RPA.

Some of the parameters in the present calculations differ
from those used in the previous MF/RPA-model. Including
$\delta U$ given by (2.21) in the calculation of the paramagnetic heat
capacity, the fit to the experimental position of the
maximum in the Schottky anomaly is improved by a slight reduction of
the zero-field value of $\Delta$ from 0.71\us meV to 0.69\us meV. This
change is accomplish by using
$$
\eqalign{
&V_{60}^{}=-3.42\,10^{-5}\hbox{meV}\cr
&V_{62}^{}=(6.41-i\,4.59)\,10^{-5}\hbox{meV}}\eqno(3.2)
$$
whereas the remaining crystal-field parameters are the same as in
Table I in Ref.1. The change also influences weakly the matrix element
($M=6.551$ instead of 6.541). The quality of the fit to the
experimental heat capacity is much improved in comparison with the one
derived in the MF-approximation, as shown in Fig.4. The fit is now
close to perfect below 7\us K, where the phonons are unimportant, and
down to about 1\us K, where critical fluctuations may start to be of
importance.

The dominant part of the two-ion coupling, due to the classical
dipole--dipole interaction, is unchanged. Given the positions of the
magnetic Ho-ions, the dipole--dipole interaction is
calculated by the Ewald method utilizing the refinements
developed by Bowden and Clark.$^{12}$ In order to get agreement with
the observed transition temperature, and to explain the larger
splitting between the $A_1^\ast$ and $A_2^{}$ excitation modes at $(1\,0\,0)$
than that predicted by the classical coupling, two exchange-coupling constants,
$\J_{13}^{}$ and $\J_{12}^{}$, were used as fitting parameters. They
are the isotropic couplings between  respectively nearest
and next-nearest neighbors. To these we here add one more
coupling, $\J_{11}^{}$, which is the nearest-neighbor coupling
between ions on the same sublattice (the distance between these ions
is not much greater than the smallest one). The final values of the
three parameters are:
$$
\eqalign{
&\J_{13}^{}=-0.30\hbox{\us $\mu$eV}\quad(-0.64\hbox{\us $\mu$eV})\cr
&\J_{12}^{}=\phantom{-}0.37\hbox{\us $\mu$eV}\quad(0.30\hbox{\us
$\mu$eV})\cr
&\J_{11}^{}=\phantom{-}0.12\hbox{\us $\mu$eV}\quad(0)}\eqno(3.3)
$$
where the numbers in the brackets are the previous values used in the
MF/RPA-model (when $\theta=+25^\circ$). The introduction of the
coupling parameter $\J_{11}^{}$ leads here to a slight improvement of the
fit to the excitation energies, and thus probably also to a
better estimate of the density of states, but it is in no way
essential for the analysis. For instance, the reduction of the
calculated energy of the upper $A_1^{}$-mode close to $(1\,0\,0)$ is a pure
renormalization effect.
The new values of the exchange parameters mean that the effective
coupling parameter determining $T_C$ (see Ref.1)
$$
\J_1^{}(\theta)=2\J_{13}^{}\cos2\theta+2\J_{12}^{}+2J_{11}^{}+16.848\J_D^{}
\eqno(3.4)
$$
is now 7.92\us$\mu$eV, which is about 11\% larger than in the MF-model.
This increase is required in order to compensate for the reduction,
by the factor $1+\Sigma(0)$, of the effective single-site susceptibility
deriving from the fluctuations. This factor is calculated to be 1.128
just above $T_C$.

In the calculation of the zero-frequency susceptibility, it is
important to include the nuclear contribution due to the hyperfine
interaction. This coupling enhances the susceptibility at $T_C$ by
about 16\%. To a first approximation the influence of the
fluctuations on this coupling may be neglected. However, it is
practicable to include the effects of the singlet--singlet fluctuations
on the hyperfine coupling, leaving out the intrinsic
$1/z$-modification of the nuclear susceptibility (which is of the
order $A^4$). In
this approximation the $xx$-component of the effective non-interacting
susceptibility, $\w{\chi}_{}^{\,o}=\chi_{}^{\,o}/\{1+\Sigma(0)\}$, is
replaced by
$$
\w{\chi}_{\rm eff}^{\,o}=\w{\chi}_{}^{\,o}(1+A_{}^2\w{\chi}_{}^{\,o}
{\chi}_{I}^{})
\eqno(3.5)
$$
and $K(0)$ by approximately $K(0)(1+A_{}^2\w{\chi}_{}^{\,o}{\chi}_{I}^{})$.
$I=7/2$ is the nuclear spin, $A=3.36$\us$\mu$eV is the hyperfine
coupling and $\chi_I^{}$ is the nuclear $xx$-susceptibility component
as determined from $\H_{\rm hf}^{}({\rm eff})$ given by (3) in Ref.1.
$\C$ in this Hamiltonian is replaced by
${1\over{N}}\sum_{\q}\B(\q)$, but
this change can be neglected in the quadrupole term.

The maximum value of the renormalization factor, $1+\Sigma(0)$, is
1.140 at $T\simeq2$\us K. Above this temperature $\Sigma(0)$ steadily
decreases, and the renormalization is reduced to 1.098 at
4.2\us K. The rather weak variation of the renormalization in this
temperature interval means that the result for the renormalized
temperature-dependent bulk susceptibility in the paramagnetic phase
is close to the previous MF-result shown in Fig.7 in Ref.1. In the
presence of a magnetic field, or in the ordered phase, the mean field
is determined by (2.33). The change of this field, $\delta H$,
induced by the fluctuations, and the moments themselves are calculated
fully self-consistently. Above $T_C$ in the low-field limit
$\delta H/H_{\rm MF}^{}=\Sigma(0)$. At higher fields,
$\delta H$ goes
through a maximum and then starts to decrease. At low temperatures the
maximum value is about 0.3\us kOe, and the maximum occurs when the
moment is about half its saturation value ($H_{\rm MF}^{}\simeq4$\us
kOe). In the high field limit $\delta H/H_{\rm MF}^{}$ vanishes,
whereas $\delta H$ itself becomes small but not zero.
It is uncertain whether this is due to the approximations made
or not, but the non-zero value of $\delta H$ at infinite fields has no
consequences. More importantly, the area determined by $\delta H$ as a
function of $g\mu_B^{}\la J_x^{}\ra$ is found to agree quite
accurately with the calculated energy change of the non-magnetic state
induced by the fluctuations. In the low-temperature limit $\delta
U/N=-9.06$\us$\mu$eV and (2.34) is satisfied to within 2--3\%. Only in
the high-temperature limit is the integral in (2.34) calculated to
vanish somewhat faster than $\delta F(m=0)$, indicating that the
elastic contributions should  be included in a more careful manner in
this limit.

At low fields the calculated renormalized magnetization is
close to that predicted by the MF-model, corresponding to the
behavior of the (zero-field) susceptibility. When the field is
applied along the $a$-direction, the magnetization increases very
rapidly and is close to its saturation value already at a field of about
10\us kOe. At this field the $1/z$-renormalization is nearly quenched,
corresponding effectively to an enhancement of the coupling
constant. The calculated magnetization at low temperatures and
intermediate fields is therefore increased in comparison with the
result derived from the MF-model. At 4.2\us K, when the field is
applied along the $c$-axis, the (change of the) renormalization effect
is small. In the other case, when the field is along the $a$-axis at
1.6\us K, the magnetization is calculated to be somewhat larger than
predicted by the MF-model, and thus closer to the experimental
behavior, as shown in Fig.5. The figure only contains the results
below 25\us kOe. The magnetization curves have been measured
up to a field of 80\us kOe. At the higher fields the results of the
two models coincide and are in good agreement with experiment (see
Fig.6 in Ref.1). The rapid quenching of the fluctuations produced by a
field along the $a$-axis at low temperatures is also illustrated by
Fig.6, showing the field dependence of the excitations observed at
$(0\,0\,1)$ at 1.6\us K. The width of the excitation band, which is
close to the distance between the upper and lower excitation in
this figure, is reduced by a factor of 5 at 10\us kOe, compared to its
zero-field value. Above about 25\us kOe the excitation band is almost
completely flat and the single-site fluctuations are of no
importance. The change of the bare interaction parameters means that
the present model reproduces the experimental behavior somewhat more
accurately than the previous model.

Just above $T_C$, the entropy due to the electronic singlet--singlet states
is vanishingly small, and the heat capacity in the ferrimagnetic phase
is determined by the variation in the population of the nuclear levels
as in the MF-case (thus the calculation of the heat capacity in the
ordered phase does not rely on the relation (2.34)).
The effects of the single-site fluctuations are included via
the reduction of the mean field derived from (2.33). The result is
compared with the experimental data in Fig.4 showing that the present
model, in contrast to the MF-model, can almost account for
the large jump in the heat capacity observed experimentally at $T_C$.
The improved description of the heat capacity in the ordered phase is
related directly to the fact
that the ordered moment predicted in the zero-temperature limit is
closer to the observed value. As shown in Fig.7, the discrepancy is
reduced by almost a factor of two, and the calculated moment at $T=0$
is now about 9\% instead of 16\% smaller than that determined$^2$ by the
neutron-diffraction experiment. The larger value of the ordered
moments implies an increase in the excitation energies and, as shown
in Fig.8, the calculated excitation energies at 0.09\us K are slightly
above those derived from the MF-model. It is argued in Ref.1 that the
sample temperature at which the experimental results shown in Fig.8
were obtained might have been higher than indicated, because  of the
very large nuclear heat capacity at these temperatures. Here we find
that we get a reasonable agreement between theory and experiment, if
the sample temperature is assumed to be about 0.25\us K.

\bigskip
\noindent
{\bf IV. Discussion and conclusion}
\medskip
\noindent
The renormalization effects due to single-site fluctuations have been
included to leading order in the case of the singlet--singlet system
HoF$_3$. The expansion of the Matsubara Green function, applicable to
this system, is considered to first order in $1/z$. In this order all
the single sites are assumed to be placed in equivalent surroundings. The
fluctuations in this effective medium, which derive self-consistently
from the response of the single sites, affect the single-site Green
function in a manner which may be determined by a cumulant expansion. An
examination of the cumulants in the series of the
single-site Green function shows that the usual `unconditional'
$1/z$-expansion could be improved substantially in a straightforward
way. One additional advantage of the present
procedure is that it is fully self-consistent, and the results
therefore also behave properly close to a second-order phase transition.

HoF$_3$ is almost the ideal system for studying the effects of
fluctuations. The electronic properties are nearly determined by
the two lowest singlets alone at low temperatures, and
the classical dipole--dipole coupling is a factor of ten larger than
other dipole couplings in the system. The single-site properties are
simple and the most important part of the two-ion interactions may be
calculated directly. Furthermore, the long-range critical
fluctuations close to the phase transition are expected to be of
marginal importance, as they only lead to logarithmic corrections to the
mean-field behavior. The effects of the $1/z$-fluctuations are calculated to
be rather substantial in this system. For instance, the single-site
susceptibility is found to be reduced effectively by up to 12\% at the lowest
temperature. In the comparison with the results of the MF-model in
Ref.1, these differences are not always manifest, because the
renormalization effects to some extent are included in this model via
an effective adjustment of the interaction parameters. The
renormalization effects vary slowly with temperature
but may be quenched rather rapidly by applying a magnetic field, in
which case differences may appear between the two models, since
the bare interaction parameters are different.

The comparison between theory and experiment has been improved
systematically by the inclusion of the effects due to the single-site
fluctuations. The most striking improvement is found in the case of
the heat capacity. Even so, the experimental heat capacity does not approach
zero as closely as the theoretical prediction at temperatures just above
$T_C$, indicating that the critical fluctuations are of some
importance, or alternatively that effects of the order
$(1/z)^2$ should be included at the lowest temperatures. Although
the theoretical magnetization curves, as a function of field or
of temperature below $T_C$, are in better agreement with
experiments than found in the MF-model, there are still some
discrepancies. We do not expect that higher-order
renormalization effects can eliminate these deviations.
An estimate of the bare electrostatic interaction between the
quadrupole moments of the 4f-electrons on the different
ions indicates that this coupling may be of some importance.
However, this interaction is expected to be strongly shielded
by the other electrons on the Ho- and the Cl-ions, and the good
account of the heat capacity does not leave much room for
any additional couplings. The discrepancies are systematic, but
they are approaching a level where experimental uncertainties may be
significant for the comparisons. Utilizing the correlation between
the jump in the heat capacity at $T_C$ and the zero-temperature
moment, the comparison of experiment with theory suggests that this moment
should be only a few per cent larger than calculated, between
5.2--5.4\us$\mu_B^{}$, to be compared with the neutron
diffraction result$^2$ of $5.7$\us$\mu_B^{}\pm 0.2$\us$\mu_B^{}$.
It is therefore apparent that a further refinement of the theoretical
understanding of this unique magnetic system must depend on the
performance of even more precise experiments.
\medskip
\noindent
{\bf Acknowledgements}
\smallskip
\noindent
The author would like to thank A.R. Mackintosh for a number of useful
comments.
\vfill\eject
\hbox{}
\bigskip
\noindent
{{\bf REFERENCES}
\parskip 8pt
\medskip

\item{$^1$}M.J.M. Leask, M.R. Wells, R.C.C. Ward, S.M. Hayden
and J. Jensen, J.\ Phys.\ Condens.\ Matter (1993).

\item{$^2$}P.J. Brown, J.B. Forsyth, P.C. Hansen, M.J.M. Leask, R.C.C. Ward
and M.R. Wells, J.\ Phys.\ Condens.\ Matter {\bf2}, 4471 (1990).

\item{$^3$}B. Bleaney, J.F. Gregg, R.W. Hill, M. Lazzouni, M.J.M. Leask
and M.R. Wells, J.\ Phys.\ C {\bf21}, 2721 (1988).

\item{$^4$}K.K. Sharma, F.H. Spedding and D.R. Blinde, Phys.\ Rev.\
{\bf24}, 82 (1981); K. Ram and K.K. Sharma, J.\ Phys.\ C
{\bf18}, 619 (1985).

\item{$^5$}J. Jensen, J.\ Phys. C {\bf17}, 5367 (1984).

\item{$^6$}R.B. Stinchcombe, J.\ Phys. C {\bf6}, 2459 (1973).

\item{$^7$}D.H.-Y. Yang and Y.-L. Wang, Phys.\ Rev.\ B {\bf10}, 4714
(1974).

\item{$^8$}C.M. Care and J.W. Tucker, J.\ Phys.\ C {\bf10}, 2773 (1977).

\item{$^9$} Y. Galili and V. Zevin, J.\ Phys.\ C {\bf20}, 2543 (1987).

\item{$^{10}$}J. Jensen and A.R. Mackintosh, {\it Rare Earth
Magnetism: Structures and Excitations} (Oxford University Press,
Oxford, 1991).

\item{$^{11}$}J. Jensen, K.A. McEwen and W.G. Stirling, Phys.\ Rev.\ B
{\bf35}, 3327 (1987).

\item{$^{12}$}G.J. Bowden and R.G. Clark, J.\ Phys.\ C {\bf14}, L827
(1981).
\par}

\vfill\eject
\hbox{}
\bigskip
\bigskip
\noindent
{\bf FIGURE CAPTIONS}
{\parskip 10pt
\parindent 30pt
\bigskip

\item{Figure 1} The 4 Ho- and 12 F-ions in one unit cell of the
orthorhombic structure of HoF$_3$,  projected on the $a$--$c$ and the
$a$--$b$ plane. The arrows indicate the local
$x$-axes lying in the $a$--$c$ plane and making the angles $\pm\theta$
with the $a$-axis, which are also the directions of
the moments in the ordered phase. The figure shows the case where
$\theta=+25^\circ$.

\item{Figure 2} The final calculated density of states per Ho-ion in the
low-temperature limit of the paramagnetic phase, ${\cal
N}(\varepsilon)$. The square-root singularities at the minimum and
maximum energies are modified by the directional dependence of the
long-wavelength $A_1$-mode (see Fig.3).

\item{Figure 3} The dispersion relation of the singlet--singlet
excitations along {\bf a}$^\ast$ and {\bf c}$^\ast$ in
the paramagnetic phase  of HoF$_3$ at 1.6\us K. The closed circles are
the experimental results$^1$ obtained with the neutron-scattering vector
along $(h\,0\,0)$ and $(0\,0\,l)$, whereas the open circles are the
results obtained along the energetically-equivalent directions
$(h\,0\,1)$ and $(1\,0\,l)$ (with $h$ and $l$ lying between 0 and 1). The
results are shown in a double zone representation, and the thin dashed lines
indicate the Brillouin-zone boundaries. The solid lines are the
theoretical predictions including the single-site fluctuations,
and the thick dashed lines are the RPA-results derived from the MF-model in
Ref.1. The labeling of the different modes, $A_1$ -- $A_4$, close to the
Bragg points is explained in Ref.1. The short solid
lines marked $A_1^\ast$ and $A_3^\ast$ show the energies of the two
ferromagnetic modes in the long wavelength limit, when the direction of
the wave-vector is perpendicular to respectively $(h\,0\,0)$ and
$(0\,0\,l)$. In combination with the non-zero experimental resolution width,
the $A_1^\ast$ mode in particular contributes much more strongly to the
scattering cross-section than the $A_1^{}$ mode close to $(1\,0\,0)$.
Resolution effects combined with the singular behavior of the long
wavelength $A_1^{}$-mode may also be important for explaining the
difference between the theoretical and experimental energy of this mode.

\item{Figure 4} The low-temperature heat capacity of HoF$_3$. The
solid circles are the experimental results obtained by Bleaney
\et,$^3$  the dashed line is the magnetic contribution predicted by
the MF-theory,$^1$ and the solid line is the result derived in the
present approximation including the effect of the single-site
fluctuations. Above 7--8\us K the phonons start to contribute, as do
the higher-lying 4f-levels on the Ho-ions.

\item{Figure 5} The magnetic moment of HoF$_3$ as a function of field
along the $a$-axis at 1.6\us K and along the $c$-axis at 4.2\us K. The
experimental results of Bleaney \et$^3$ have been corrected for the
demagnetization field estimated from the shape of the samples. The
mean-field predictions$^1$ are shown by the dashed lines, and the
present calculations by the solid lines.

\item{Figure 6} The position of the inelastic scattering peaks at
$(0\,0\,1)$ as a function of magnetic field applied along the
$a$-direction at 1.6\us K. The circles are the experimental results$^1$
and the dashed and solid lines show the calculated energies of the $A_2$,
$A_3^\ast$ and $A_3$ modes, respectively, in the RPA$^1$ and when
including the $1/z$-renormalization. The $A_2$ and $A_3$ modes are near the
minimum and maximum excitation energies, and the decreasing distance
between the upper and lower lines indicates that the total excitation
bandwidth rapidly declines as the field is increased.

\item{Figure 7} The solid lines show the calculated values of the
angular moment, $\la J_x\ra$, and the nuclear spin, $\la I_x\ra$, of a
Ho-ion as  a function of temperature below $T_C$. The dashed lines are
the corresponding results derived in the MF-case.$^1$ The filled circles are
the experimental values of $\la J_x\ra$ determined from the variation
of the magnetic scattering intensities at $(1\,0\,0)$ obtained by by
Brown \et.$^2$ Their results have been scaled to agree with the
magnetic moment of 5.7\us$\mu_B^{}$, which they obtained from structure
refinements at 70\us mK.

\item{Figure 8} The dispersion relation along {\bf a}$^\ast$ and
{\bf c}$^\ast$
in the ordered phase of HoF$_3$ at 90\us mK. The meaning of
the symbols is the same as in Fig.3. The cross-hatched lines show the
theoretical results at 0.25\us K.\par
}
\vfill\eject
}\end